\documentclass[preprint]{aastex63}

\usepackage{hyperref}

\received{August 7, 2020}
\revised{September 7, 2020}
\accepted{September 23, 2020}
\submitjournal{ApJ}

\shorttitle{Hemispheric Sign Preference of Magnetic Helicity Flux}
\shortauthors{Park et al.}

\graphicspath{{./}{figures/}}

\begin{document}

\title{Magnetic Helicity Flux across Solar Active Region Photospheres: I. Hemispheric Sign Preference in Solar Cycle 24}

\correspondingauthor{Sung-Hong Park}
\email{shpark@isee.nagoya-u.ac.jp}

\author[0000-0001-9149-6547]{Sung-Hong Park}
\affiliation{Institute for Space-Earth Environmental Research, Nagoya University, Nagoya, Japan}

\author[0000-0003-0026-931X]{K. D. Leka}
\affiliation{Institute for Space-Earth Environmental Research, Nagoya University, Nagoya, Japan}
\affiliation{NorthWest Research Associates, Boulder, CO, USA}

\author[0000-0002-6814-6810]{Kanya Kusano}
\affiliation{Institute for Space-Earth Environmental Research, Nagoya University, Nagoya, Japan}

\begin{abstract}
A hemispheric preference in the dominant sign of magnetic helicity has been observed in numerous features in the solar atmosphere: i.e., left-handed/right-handed helicity in the northern/southern hemisphere. The relative importance of different physical processes which may contribute to the observed hemispheric sign preference (HSP) of magnetic helicity is still under debate. Here, we estimate magnetic helicity flux ($dH/dt$) across the photospheric surface for 4,802 samples of 1,105 unique active regions (ARs) that appeared over an 8-year period from 2010 to 2017 during solar cycle 24, using photospheric vector magnetic field observations by the Helioseismic and Magnetic Imager (HMI) onboard the Solar Dynamics Observatory (SDO). The estimates of $dH/dt$ show that 63\% and 65\% of the investigated AR samples in the northern and southern hemispheres, respectively, follow the HSP. We also find a trend that the HSP of $dH/dt$ increases from $\sim$50\,--\,60\% up to $\sim$70\,--\,80\% as ARs (1) appear at the earlier inclining phase of the solar cycle or higher latitudes; (2) have larger values of $|dH/dt|$, the total unsigned magnetic flux, and the average plasma flow speed. These observational findings support the enhancement of the HSP mainly by the Coriolis force acting on a buoyantly rising and expanding flux tube through the turbulent convection zone. In addition, the differential rotation on the solar surface as well as the tachocline $\alpha$-effect of flux-transport dynamo may reinforce the HSP for ARs at higher latitudes.
\end{abstract}


\section{Introduction} \label{sec:intro}
Over the past decades, extensive observations of the solar atmosphere from the photosphere to the corona have revealed a hemispheric preference in the dominant sign of magnetic helicity, specifically a dominance of negative (left-handed) helicity in the northern hemisphere and positive (right-handed) helicity in the southern hemisphere, independent of the solar cycle. This is the so-called hemispheric sign preference (hereafter referred to as HSP) observed in numerous magnetic structures in the Sun. For example, a majority of sunspot penumbral fibrils in the northern/southern hemisphere show counterclockwise/clockwise rotations from outside to inside \citep{1927Natur.119..708H,1941ApJ....93...24R,1987SoPh..107..221D,2003AdSpR..32.1867P}. Solar prominences in the northern/southern hemisphere tend to be composed of dextral/sinistral channels with right-bearing/left-bearing barbs \citep{1994ASIC..433..303M,2003ApJ...595..500P,2009ApJ...692..104L,2017ApJ...835...94O}; this tendency is found in all cases of prominences and filaments: active-region, intermediate, and quiescent, with different degrees of compliance. Bright coronal loops displaying a distinct ``S-shape'' (called sigmoids) show their hemispheric preference in their shape as observed in (for example) soft X-ray emission, i.e., an inverse/forward S-shape in the northern/southern hemisphere \citep{1996ApJ...464L.199R,1999GeoRL..26..627C,2014SoPh..289.3297S}. This HSP is also found in strong-field plage and sunspot regions, as well as in quiet-Sun network regions, by means of the force-free-field parameter $\alpha$ and the vertical component of current helicity density calculated from photospheric vector magnetograms \citep{1995ApJ...440L.109P,1999ApJ...519..876Z,2001ASPC..236..423P,2013MNRAS.433.1648G,2013ApJ...772...52G,2015PASJ...67....6O}. Furthermore, near-Earth {\it in situ} observations of interplanetary coronal mass ejections (ICMEs) show that the handedness of twisted magnetic field lines in ICMEs is well matched with that estimated in their solar source ARs \citep{2013SoPh..284..105C}, exhibiting the relevant HSP.

Several physical processes have been proposed to explain the observed HSP in these solar magnetic structures \citep[refer to a review by][and references therein]{2002ApJ...573..445B}. The proposed processes are thought to play a crucial role in the generation of twist on magnetic flux tubes at the base of the convection zone, during their buoyant rise through the bulk of the convection zone, and by their footpoint motions in the photosphere, such as: (1) differential rotation \citep{1990SoPh..125..219S,2000JGR...10510481B,2000ApJ...539..944D,2002SoPh..207...87D}, (2) Coriolis force \citep{1985ApJ...291..300G,1997ApJ...488..443L,2013ApJ...775L..46W}, (3) $\alpha$-effect involved in turbulent dynamo models \citep{1996GeoRL..23.2649R,1996PhRvE..53.1283S,1999GMS...111...75G,2001ApJ...559..428D,2013ApJ...768...46P}, (4) helical turbulent convection \citep[called $\Sigma$-effect;][]{1998ApJ...507..417L}, and (5) local photospheric flows associated with magnetic flux cancellation and reconnection \citep{1989ApJ...343..971V,1994ApJ...427..459P,1999ApJ...515..435L,1999GMS...111..213V}. As discussed in \citet{2002ApJ...573..445B}, some of the aforementioned processes need not be mutually exclusive but may simultaneously operate in producing twist on flux tubes through the layers from the deep convection zone to the photosphere. As a result, the degree of the HSP may be either strengthened or weakened by multiple combinations of the processes acting on rising flux tubes. However, the relative importance of these different processes responsible for the HSP is still under debate. It is also not clear what causes variations in the strength of the HSP, which were observed, for example, in different phases of a solar cycle \citep{2005PASJ...57..481H,2019ApJ...882...80G,2019ApJ...887..192K} as well as in different solar features \citep{2014SSRv..186..285P}.

Magnetic helicity is a useful, quantitative measure to diagnose the underlying physical processes that govern the HSP, in terms of twists, kinks, and inter-linkages of magnetic field lines \citep[e.g.,][]{1984JFM...147..133B,2006JPhA...39.8321B,2009AdSpR..43.1013D}. Here we estimate magnetic helicity flux across the photospheric surface of active regions (ARs) over an 8-year span of 2010 through 2017, using photospheric vector magnetic field observations by the Helioseismic and Magnetic Imager \citep[HMI,][]{2012SoPh..275..207S} onboard the Solar Dynamics Observatory \citep[SDO;][]{2012SoPh..275....3P}. Analyzing the estimated magnetic helicity flux with a large sample, we intend to (1) quantitatively investigate the HSP of magnetic helicity flux for different groups as categorized by their heliographic latitudes and phase of the solar cycle, Mount Wilson magnetic classifications, and relevant magnetic field properties, (2) investigate if there are trends in the strength of the HSP with respect to those considered categorizations, and (3) support (or not) a preferred condition and relevant mechanism for adherence to the observed HSP.

\section{Observations and Data Analysis} \label{sec:obs}
\subsection{Data Set of Active Region Vector Magnetograms} \label{subsec:obs_dataset}
In this study, we analyze a data set of HMI full-disk photospheric vector magnetograms ({\tt hmi.B\_720s}) with a spatial resolution of 1{\arcsec} (pixel size of 0.5{\arcsec}) observed at 00:36 and 00:48 TAI each day from 2010 May 1 to 2017 December 31. For the given HMI data set, we first search for automatically identified HMI AR Patches \citep[HARPs;][as recorded in the {\tt hmi.M\_720s} series]{2014SoPh..289.3483H}, each of which satisfies the following criteria: (1) the HARP boundaries are located within $\pm$60$^\circ$ in longitude from the central meridian, and (2) only a single NOAA-numbered AR with at least one sunspot visible in white-light is contained in the HARP field of view (FOV). Hereafter, we will refer to such HARPs satisfying our selection criteria as ARs. HARPs containing a single NOAA-numbered AR but without a sunspot are not considered as ARs, but are categorized instead as a distinct group, namely old-region plage. Because NOAA assigns numbers only to active regions containing sunspots, numbered regions without sunspots can uniquely be considered old, decaying active region remnants. A comparison of the HSP of magnetic helicity flux between ARs and plages is presented in the Appendix.

\begin{figure}[t!]
\centering
\includegraphics[width=0.88\textwidth]{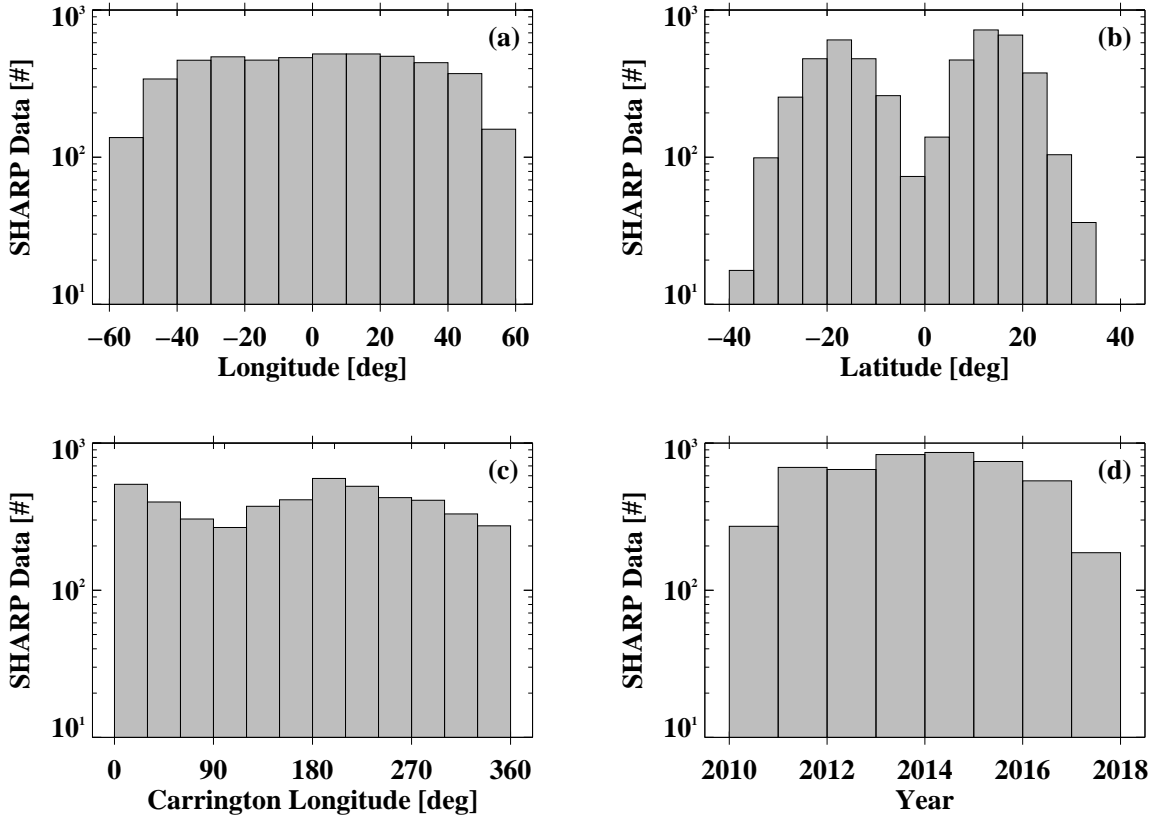}
\caption{Frequency distributions of 4,802 HARPs considered in this study for estimation of magnetic helicity flux, with respect to (a) heliographic longitudes, (b) heliographic latitudes, (c) Carrington longitudes, and (d) observation years of the HARPs.}
\label{fig:harp_dist}
\end{figure}

The coordinates of each identified HARP are used to construct cutout images of all three vector magnetic field components from the corresponding full-disk vector magnetic field data, using the NWRA remote-SUMS/DRMS system. Over the examined period of 2010 to 2017, consequently, we obtain a total of 4,802 co-aligned pairs of AR vector magnetograms with time separation of 12 minutes, which we consider 4,802 samples. Given the definition above, this amounts to 1,105 unique NOAA numbers. While many ARs are thus sampled repeatedly with a 24 hr cadence, we consider these samples independent because magnetic helicity flux estimates can show significant variations, including sign changes, over timescales of 24 hr and longer \citep{2008ApJ...686.1397P,2010ApJ...718...43P,2012ApJ...761..105L,2017A&A...597A.104V,2018ApJ...865..139B}. For each HARP in our data set, we examined the fraction of magnetic helicity flux measurements
that are opposite that HARP's majority sign over the observing time between the longitude limits imposed.
Upon examination, the fraction is found to be 0.17, on average, for all HARPs studied here. Such observed sign changes could be caused by difference processes such as emerging magnetic flux with an opposite sign of magnetic helicity into the corona, submerging magnetic flux into the interior that brings magnetic helicity across the surface, and photospheric flows that unwind twisted magnetic field lines. Of note, because we are sampling the HMI data at the same time each day, variations in the magnetic helicity flux are likely not due to the influence of the spacecraft orbital velocity \citep{2012ApJ...761..105L,2016ApJ...823..101S}. Figure~\ref{fig:harp_dist} shows the frequency distributions of (a--c) center coordinates, and (d) observed years of all identified HARPs at 00:36 TAI. All the co-aligned AR vector magnetogram pairs are remapped from pixel coordinates in the original CCD image plane to helioplanar coordinates (here, denoted as $B_{x}$, $B_{y}$ and $B_{z}$). The $x$-axis and $y$-axis are oriented toward solar west and north, respectively, on a plane tangent to the solar surface at a point at the center of the HARP FOV, and the $z$-axis is in the vertical direction to the $x$--$y$ plane, following \citet{1990SoPh..126...21A}.

\subsection{Estimation of Photospheric Magnetic Helicity Flux} \label{subsec:obs_dhdt}
Photospheric plasma velocity fields (i.e., $v_{x}$, $v_{y}$ and $v_{z}$) in ARs are derived from the co-aligned pairs of $B_{x}$, $B_{y}$ and $B_{z}$ images at 12-minute separation with the Differential Affine Velocity Estimator for Vector Magnetograms \citep[DAVE4VM;][]{2008ApJ...683.1134S}. 
The apodization window size used in DAVE4VM is set to 13\,pixels, based on the same evaluation test performed by \citet{2008ApJ...683.1134S}, but for the pairs of HMI $B_{x}$, $B_{y}$ and $B_{z}$ images in our case.

With the AR photospheric magnetic and velocity fields, we calculate the net magnetic helicity transfer rate $dH/dt$, also known as magnetic helicity flux, across the entire photospheric surface $S$ (i.e., the HARP FOV) of each AR
\begin{equation}
\frac{dH}{dt} = \int_{S} G_{\theta} (\textit{\textbf{x}}) \, dS, \label{eq:dhdt}
\end{equation}
where $G_{\theta}(\textit{\textbf{x}})$ is a photospheric surface density of magnetic helicity flux given in \citet{2005A&A...439.1191P} as
\begin{equation}
G_{\theta} (\textit{\textbf{x}}) = -{\frac{B_{z}}{2\pi}} \int_{S^\prime} \frac{\hat{\textit{\textbf{z}}} \cdot \left((\textit{\textbf{x}}-\textit{\textbf{x}}^\prime) \times (\textit{\textbf{u}}-\textit{\textbf{u}}^\prime)\right)} {{|\textit{\textbf{x}}-\textit{\textbf{x}}^\prime|^2}} B^\prime_{z} \, dS^\prime.
\label{eq:g_theta}
\end{equation}
As described in \citet{2003SoPh..215..203D}, here the photospheric flux transport velocity $\textit{\textbf{u}}$ is the horizontal velocity of photospheric footpoints of magnetic field lines defined as
\begin{equation}
\textit{\textbf{u}} = \textit{\textbf{v}}_{h} - \frac{v_{z}}{B_{z}}\textit{\textbf{B}}_{h},
\label{eq:footpoint_vel}
\end{equation}
where the subscript ``$h$'' represents the horizontal components of photospheric vector fields. It should be noted that the surface integrals in Equations~(\ref{eq:dhdt}) and~(\ref{eq:g_theta}) are carried out only for high-confidence pixels (as indicated by a value of 90 in the {\tt conf\_disambig.fits} segment; see \citep{2014SoPh..289.3483H}). Throughout this paper for notational simplicity, magnetic helicity refers to gauge-invariant relative magnetic helicity \citep{1984JFM...147..133B,1985CoPPC..9..111} with the specific choice of the Coulomb gauge for the vector potential of the potential field. $dH/dt$ reflects only magnetic helicity flux across the photosphere for a region of interest, such as would be caused by magnetic helicity transport due to emerging/submerging twisted magnetic flux tubes or photospheric footpoint motions. The estimated $dH/dt$ in this study therefore does not necessarily represent the HSP of the overall magnetic helicity content in the full coronal volume above said region of interest.

\begin{figure}[t!]
\centering
\includegraphics[width=0.92\textwidth]{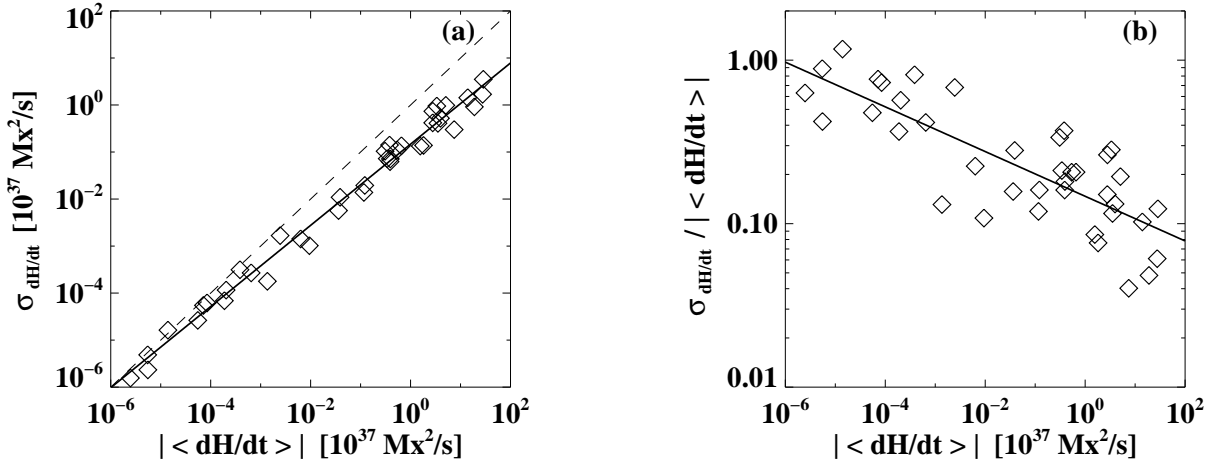}
\caption{Results from Monte Carlo simulations to determine the uncertainty in the calculation of $dH/dt$ with a sample of 39 unique NOAA-numbered ARs. Scatter plots of (a) the standard deviation ($\sigma_{dH/dt}$) and (b) the ratio of $\sigma_{dH/dt}$ to the absolute value of the mean ($|\!\!<\!\!dH/dt\!\!>\!\!|$) are shown as a function of $|\!\!<\!\!dH/dt\!\!>\!\!|$. The sold line in each panel represents the least squares regression line. In panel (a), the dotted line indicates the $x$\,$=$\,$y$ line.}
\label{fig:dhdt_error}
\end{figure}

\subsection{Uncertainties in Magnetic Helicity Flux Estimates} \label{subsec:obs_uncertainty}
The uncertainty in the calculation of $dH/dt$ is estimated from a Monte Carlo simulation as follows. (1) To each data point in a co-aligned pair of 12-minute-separated $B_{x}$, $B_{y}$ and $B_{z}$ maps of a given AR, we add a noise value randomly taken from a Gaussian distribution with the standard deviation corresponding to the uncertainty in each component of the magnetic field vector as propagated from the reported ``{\tt error}'' segments provided from the Milne-Eddington inversion of the HMI Stokes profiles \citep{2014SoPh..289.3483H}, assuming the disambiguation is perfect. (2) $v_{x}$, $v_{y}$ and $v_{z}$ are determined, applying the DAVE4VM method to the pair of $B_{x}$, $B_{y}$ and $B_{z}$ with random Gaussian noise added. (3) $dH/dt$ is estimated from Equations~(\ref{eq:dhdt})--(\ref{eq:footpoint_vel}). (4) The procedure of (1) to (3) is repeated 100 times so that the mean ($<\!\!dH/dt\!\!>$) and standard deviation ($\sigma_{dH/dt}$) values of $dH/dt$ are determined. Figure~\ref{fig:dhdt_error} presents the Monte Carlo simulation results for 39 different NOAA-numbered ARs. The ratio of $\sigma_{dH/dt}$ to the absolute value of the mean $|\!\!<\!\!dH/dt\!\!>\!\!|$ is considered here as the relative error in the magnitude of the estimated $dH/dt$. We find that the smaller the value of $|\!\!<\!\!dH/dt\!\!>\!\!|$, the larger the relative error. It is found that as $|\!\!<\!\!dH/dt\!\!>\!\!|$ decreases to $10^{31}$\,Mx$^{2}$\,s$^{-1}$, the relative error becomes close to unity, in which case there may be considerable uncertainties associated with the determination of the sign of $dH/dt$. Note that only $\sim$0.8\% of our examined AR vector magnetogram data set (i.e., 40 out of 4,802) show $|dH/dt|$\,$\leq$\,$10^{31}$\,Mx$^{2}$\,s$^{-1}$ so that the results from this HSP study of $dH/dt$ are thought to be minimally affected by uncertainties in magnetic helicity flux estimates.

\section{Results} \label{sec:results}
\subsection{Dependence of the Hemispheric Helicity Sign Preference} \label{subsec:results_hsp}

\begin{figure}[t!]
\centering
\includegraphics[width=0.8\textwidth]{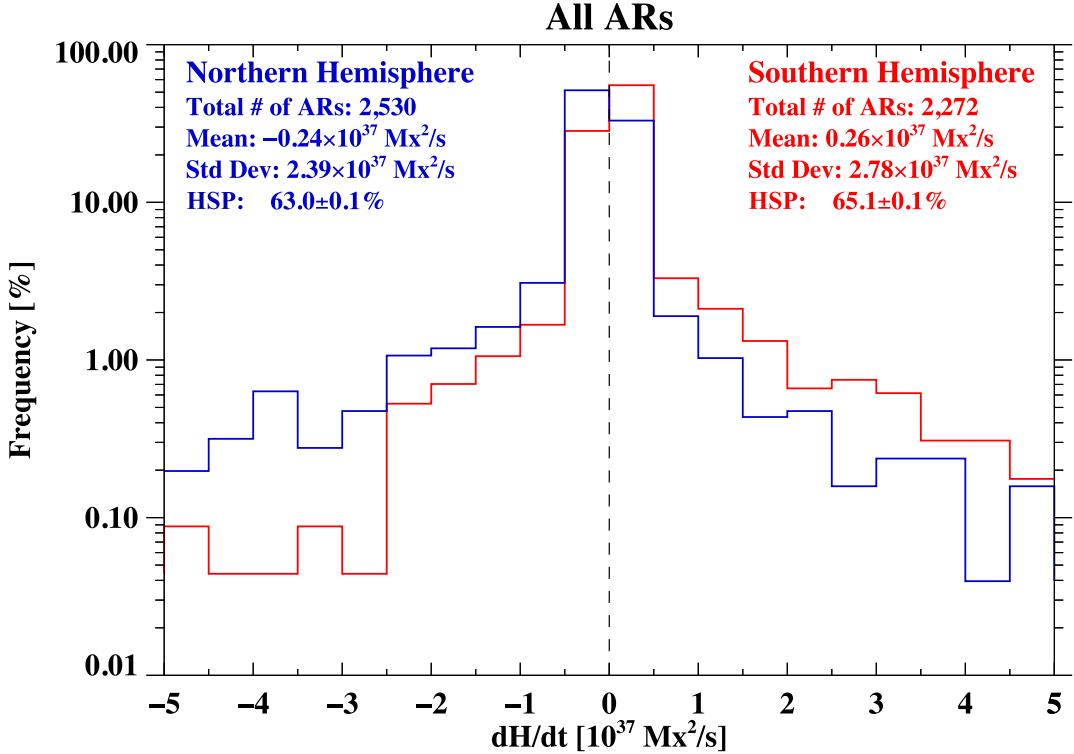}
\caption{Relative frequency distribution of $dH/dt$ values for ARs in the northern (blue) and southern hemispheres (red), respectively.}
\label{fig:hist_dhdt}
\end{figure}

\begin{figure}[t!]
\centering
\includegraphics[width=0.78\textwidth]{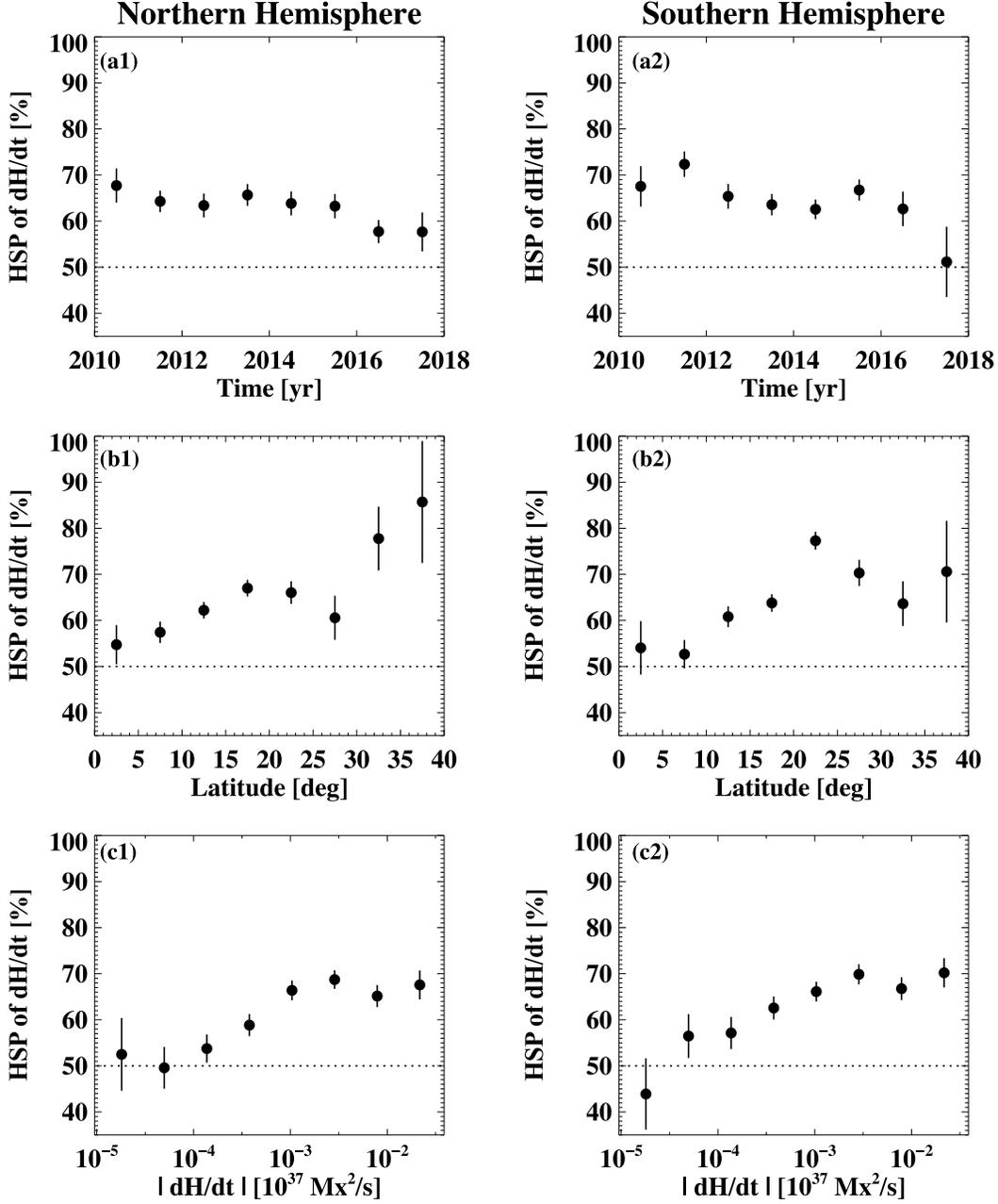}
\caption{Fraction of ARs in the northern (left column) and southern (right column) hemispheres, respectively, following the HSP of $dH/dt$, as a function of (a) date, (b) heliographic latitude, and (c) $|dH/dt|$. Error bars represent Poisson uncertainties in the HSP, each of which is calculated as $1/\sqrt{N}$, where $N$ is the total number of SHARP images pairs in each bin.}
\label{fig:hsp_1_ar}
\end{figure}

\begin{figure}[t!]
\centering
\includegraphics[width=0.78\textwidth]{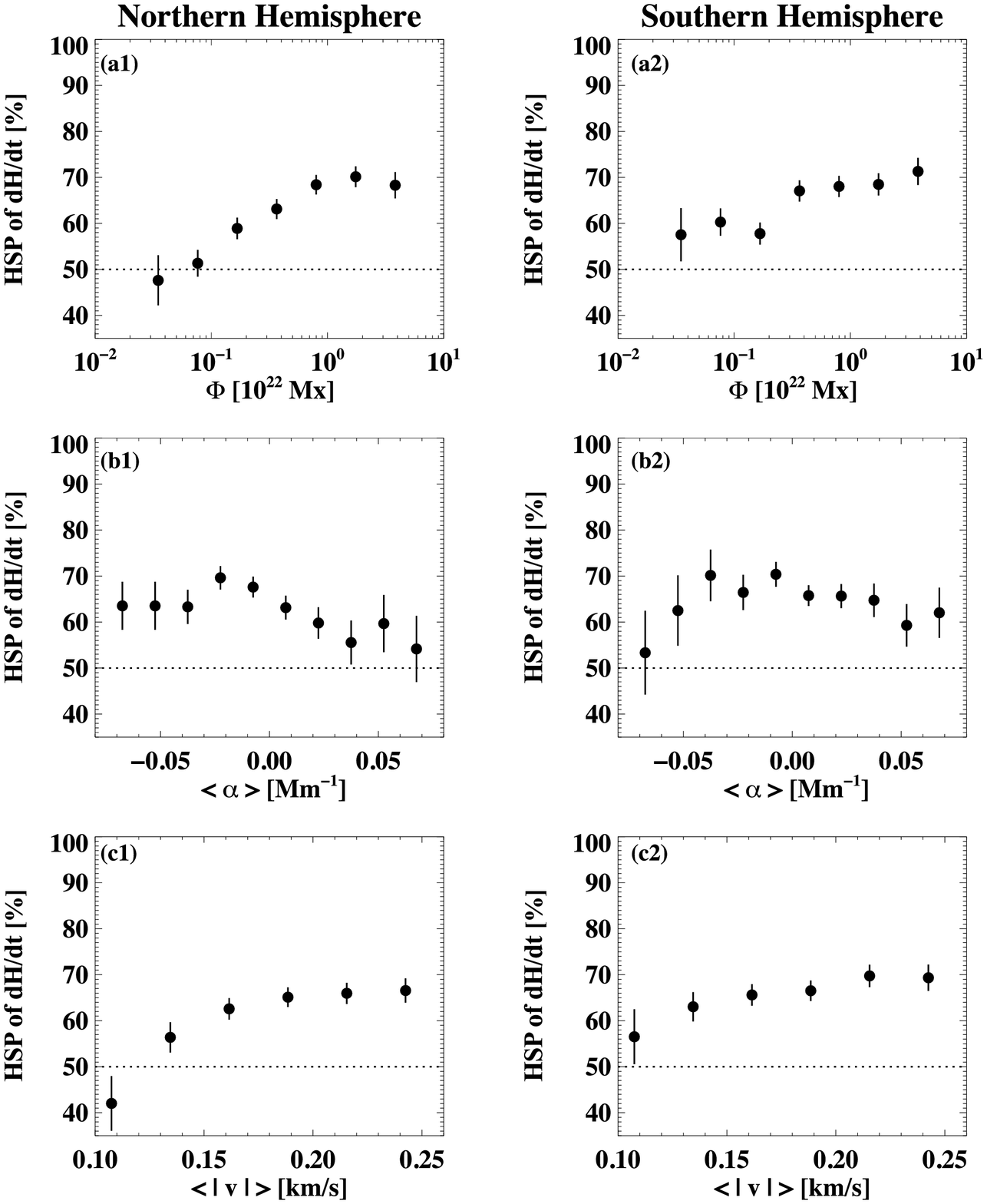}
\caption{Same as Figure~\ref{fig:hsp_1_ar}, but for the HSP of $dH/dt$ as a function of (a) the total unsigned magnetic flux $\Phi$, (b) the average force-free parameter $<\!\!\alpha\!\!>$, and (c) the average flow speed $<\!\!|v|\!\!>$ across the entire photospheric surface $S$ of the given AR.}
\label{fig:hsp_2_ar}
\end{figure}

With $dH/dt$ determined for each of 4,802 photospheric vector magnetogram pairs, we examine the degree and characteristics of the HSP over the 8-year period from 2010 to 2017 of solar cycle 24. Figure~\ref{fig:hist_dhdt} shows the relative frequency distribution of $dH/dt$ values for ARs in the northern (blue curve) and southern hemispheres (red curve), respectively. The $dH/dt$ distribution for the AR samples in the northern hemisphere exhibits a negative skewness with the mean of $-0.24\times10^{37}$\,Mx$^{2}$\,s$^{-1}$, while samples in the southern hemisphere show a positive skewness with the mean of $0.26\times10^{37}$\,Mx$^{2}$\,s$^{-1}$. 
For all of the AR samples, the degree of the HSP of $dH/dt$ is found to be 63$\pm$0.1\% of 2,530 in the northern hemisphere and 65$\pm$0.1\% of 2,272 in the southern hemisphere, which is within the range reported in previous studies \citep[e.g.,][]{1998ApJ...507..417L,2001ApJ...549L.261P,2005PASJ...57..481H,2006ApJ...646L..85Z,2014ApJ...783L...1L}. There is also a weak tendency for ARs with larger values of $|dH/dt|$ to show a higher degree of the HSP: for example, 69$\pm$0.2\% and 70$\pm$0.2\% for a subset of ARs with $|dH/dt|$\,$>$\,$10^{37}$\,Mx$^{2}$\,s$^{-1}$ in the northern and southern hemispheres, respectively. Moreover, we divided all ARs into three subgroups as to their Mount Wilson classifications (specifically the alpha, beta, and other complex classes), and include the distributions of $dH/dt$ values for each. No significant difference is however found in the distributions between the AR subgroups. Note that by analyzing the force-free parameter $\alpha$ in 148 ARs, \citet{2014ApJ...783L...1L} found a weak tendency for the alpha- or beta-class ARs to have a higher HSP than the other complex-class ARs; however, the differences in the HSP of $\alpha$ between the different Mount Wilson classifications was not statistically significant due to the small sample size.

We now investigate any dependence of the HSP of $dH/dt$ on broad properties of ARs. Figure~\ref{fig:hsp_1_ar} shows the fraction of ARs in the northern and southern hemispheres, respectively, that follow the HSP, as a function of (a) date (as a proxy for solar cycle phase, assuming a rough 2010 start date), (b) heliographic latitude, and (c) $|dH/dt|$. The HSP tends to gradually increase from $\sim$50\,--\,60\% up to $\sim$70\,--\,80\% for AR samples appearing at an earlier date of the examined period, or a higher latitude. This implies that ARs in the inclining phase of solar cycle 24 are more likely to comply with the HSP than those in the declining phase. The observed dependence of the HSP on the solar cycle phase was also found by other studies \citep[e.g.,][]{2005PASJ...57..481H,2013MNRAS.433.1648G,2014ApJ...783L...1L,2019ApJ...877L..36P} in which the HSP was evaluated with different measures of magnetic twist such as the force-free parameter $\alpha$ and vertical current helicity density. In the case of \citet{2005PASJ...57..481H}, we draw particular attention to the years categorized as activity minimum and maximum periods of solar cycles 21--23 in their study, and found that the activity minimum period mostly consists of the late declining phase of the solar cycles (i.e., several years after the sunspot maximum), while the maximum period contains the early inclining phase. We note that our study quantitatively shows the HSP of $dH/dt$ as a function of date and latitude, for the first time, from the inclining to declining phases of solar cycle 24. In addition, a weak tendency is found that ARs with larger values of $|dH/dt|$ have a higher HSP.

The HSP dependence is also examined in the context of other physical properties of AR photospheric magnetic and velocity fields. Figure~\ref{fig:hsp_2_ar} shows the HSP of $dH/dt$, as a function of (a) the total unsigned magnetic flux $\Phi$, (b) the average force-free parameter $<\!\!\alpha\!\!>$, and (c) the average flow speed $<\!\!|v|\!\!>$ through the entire HARP FOV, where $|v|=\sqrt{v_{x}^2+v_{y}^2+v_{z}^2}$. Note that $<\!\!\alpha\!\!>$ and $<\!\!|v|\!\!>$ are calculated only from the same high-confidence pixels with {\tt conf\_disambig}\,$=$\,90 as used in calculating $dH/dt$. We find a higher HSP for ARs with larger values of $\Phi$ or $<\!\!|v|\!\!>$. A similar increasing trend of the HSP with stronger magnetic fields has been reported by \citet{2020SoPh..295...106} in which they examined $\alpha$, vertical current helicity density, and kinetic helicity density parameters from measurements of photospheric magnetic fields and near surface sub-photospheric flow fields in 189 ARs observed during the descending phase of the solar cycle 23. Investigating current helicity density derived from synoptic maps of the photospheric vector magnetic field, \citet{2006ApJ...646L..85Z} and \citet{2013ApJ...772...52G} also reported that strong-field ($B_{z}$\,$>$\,1000 G) regions show a higher degree of the HSP as compared to weak-field (100 G\,$<$\,$B_{z}$\,$<$\,500 G) regions. Finally, as expected, the HSP compliance of $dH/dt$ is lower for ARs for which the sign of $<\!\!\alpha\!\!>$ is positive/negative in the northern/southern hemisphere, (i.e., opposite to that expected from the HSP, and confirming the correlation between $dH/dt$ and $<\!\!\alpha\!\!>$). 

\subsection{Interpretations} \label{subsec:results_interpret}
We offer some interpretations of our observational findings on the HSP dependence of $dH/dt$ with respect to various properties of ARs. Before interpreting the HSP dependence, we remind the reader that solar dynamo models need to reproduce these observed HSP dependencies as well as the degree of overall conforming to the HSP at the $\sim$60\% level as shown in Figure~\ref{fig:hist_dhdt}. 

First, an enhancement of the HSP can be made by the Coriolis force (including the small-scale $\Sigma$-effect) acting on a buoyantly rising and expanding flux tube through the turbulent convection zone, as it generates the twist of magnetic field lines inside the flux tube which has a sign agreeing with the HSP. The twist will be more effectively induced by the Coriolis force in cases where a flux tube (1) rises through the bulk of the convection zone at a higher latitude; (2) has larger magnetic flux so that both the rising speed and expansion of the flux tube are faster due to larger magnetic buoyancy and pressure. This scenario of the HSP enhancement is in good agreement with our observational findings that the HSP of $dH/dt$ is positively correlated with heliographic latitude, $\Phi$, and $<\!\!|v|\!\!>$, respectively. On the other hand, the Coriolis force acting on the photospheric footpoints of an emerged flux tube will produce positive/negative magnetic helicity in the northern/southern hemisphere (i.e., a negative contribution to the HSP). The observed dependencies of the HSP on heliographic latitude and $<\!\!|v|\!\!>$, however, suggest that this action of the Coriolis force may be relatively weak compared to other sources of the HSP.
 
Another important process that can generate and modify twist on flux tubes is differential rotation. A negative contribution to the HSP will be made in the convection zone by a shearing action of differential rotation on a rising $\Omega$-shaped flux tube with a tilt between its leading and following legs of opposite polarities. This suggests that a flux tube rising through a localized region of the convection zone with a larger degree of differential rotation may have an opposite sign of magnetic helicity as compared to the one expected from the HSP. In the global-scale three-dimensional convective dynamo simulations by \citet{2014ApJ...789...35F,2016AdSpR..58.1497F}, it is found that the presence of strong, large-scale magnetic fields at the base of the convection zone results in a more enhanced outward Reynolds stress and consequently a larger differential rotation. Moreover, the differential rotation in the convection zone is found to be larger at a higher latitude in both simulations and helioseismic observations \citep[see a review by][and references therein]{2009LRSP....6....1H}. In contrast to the above-mentioned possible ways of weakening the HSP by the differential rotation, our results, however, show an increasing trend of the HSP of $dH/dt$ with $\Phi$, as well as with latitude, which implies that the HSP may be less impacted by the differential rotation in the convection zone. Contrary to the shearing action of differential rotation in the convection zone, the photospheric differential rotation on the solar surface will lead to the enhancement of the HSP, twisting the footpoints of emerged magnetic fields and shearing a bipolar sunspot pair of an emerged $\Omega$-shaped flux tube. The observed increase of the HSP of $dH/dt$ with increasing latitude supports the scenario that the HSP is strengthened by a larger surface differential rotation observed at a higher latitude \citep{2014AJ....148..101S,2017ApJ...836...10L}.

Meanwhile, the tachocline $\alpha$-effect of flux-transport dynamos \citep[e.g.,][]{1999GMS...111...75G,2001ApJ...559..428D} at the base of the convection zone will generate a twisted flux tube of which helicity sign follows the HSP. It is shown that the relative amplitude of the tachocline $\alpha$-effect in the latitude range of $\sim$30{\degr}--50{\degr} is much larger than that at lower latitudes. Consequently, a higher HSP is expected at higher latitudes, which is supported by the observed positive correlation between the HSP of $dH/dt$ and latitude.

\section{Summary and Conclusions} \label{sec:discussion}
Magnetic helicity is a useful, quantitative measure to examine the handedness of twisted magnetic field lines observed in various magnetic structures in the solar atmosphere. In this paper we have investigated magnetic helicity flux ($dH/dt$) across the photospheric surface for 4,802 samples of 1,105 unique active regions that appeared over an 8-year period from 2010 to 2017 of solar cycle 24. Through the analysis of the AR magnetic helicity flux estimates, we found that 63\% and 65\% of the samples in the northern and southern hemispheres, respectively, follow the hemispheric sign preference (HSP): i.e., displaying predominantly the left-handed/right-handed sign of $dH/dt$ in the northern/southern hemisphere. It is also found that the HSP compliance of $dH/dt$ increases from $\sim$50\,--\,60\% up to $\sim$70\,--\,80\% in cases where the ARs (1) appeared during the inclining phase of the solar cycle, or at higher latitudes; (2) had larger values of $|dH/dt|$, the total unsigned magnetic flux, or the average plasma-flow speed; (3) displayed the same sign between the average force-free parameter and that expected from the HSP (i.e., negative/positive in the northern/southern hemisphere).

All of these observational findings lend support to the scenario that the HSP of $dH/dt$ is enhanced primarily by the Coriolis force acting on a buoyantly rising and expanding flux tube through the turbulent convection zone. In addition, the action of differential rotation on the solar surface as well as the tachocline $\alpha$-effect of flux-transport dynamos may also reinforce the HSP for ARs observed at higher latitudes. Meanwhile, the observed HSP tendencies imply that an expected weakening of the HSP by the differential rotation in the convection zone is less significant compared to the enhancement of the HSP by the Coriolis force in the convection zone, the photospheric differential rotation, and/or the tachocline $\alpha$-effect. We urge research that employs state-of-the-art solar convective dynamo simulations to explore and use as model validation the trends put forth here, in order to judge the relative input of relevant processes to helicity transport throughout the solar cycle.

\acknowledgments
The authors would like to thank an anonymous referee for constructive comments that helped clarify aspects of this paper. The authors also thank Graham Barnes, Eric Wagner, Orion Poplawski, and the NWRA/Boulder team for the development and maintenance of the SDO remote SUMS/DRMS site and analysis servers at NWRA. The data used in this work are courtesy of the NASA/SDO and the HMI science teams. This research has made extensive use of the NASA's Astrophysics Data System (ADS) as well as the computer system of the Center for Integrated Data Science (CIDAS), Institute for Space-Earth Environmental Research (ISEE), Nagoya University. This work was partially supported by MEXT/JSPS KAKENHI Grant No.~JP15H05814.
\vspace{5mm}
\facilities{SDO (HMI)}
\software{DAVE4VM \citep{2008ApJ...683.1134S}}

\clearpage

\section*{Appendix: Comparison of the Hemispheric Helicity Sign Preference between Active Regions and Plage Regions} \label{sec:appendix}
\begin{figure}[t!]
\centering
\includegraphics[width=0.78\textwidth]{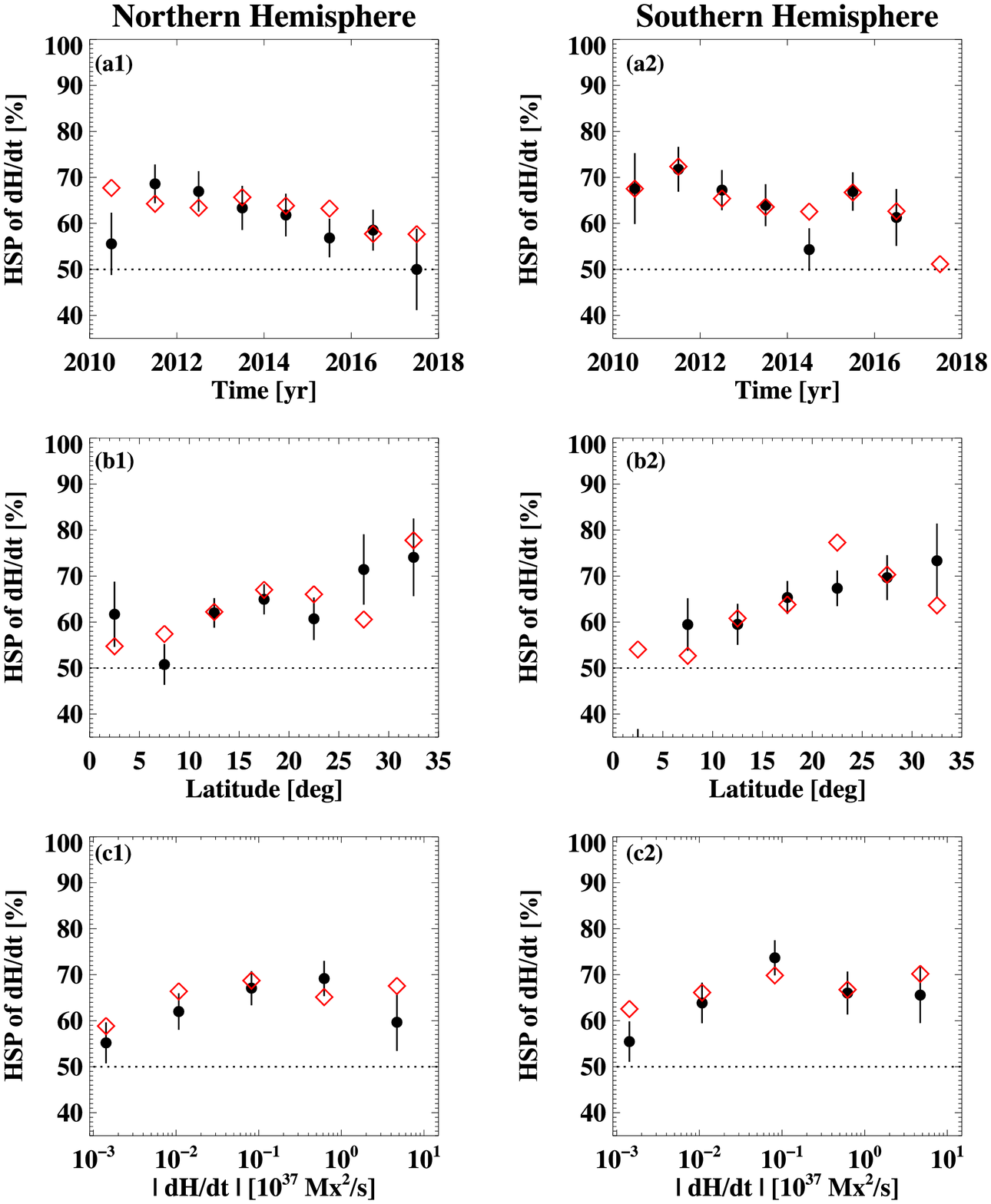}
\caption{Same as Figure~\ref{fig:hsp_1_ar}, but for plage areas (circles with error bars). The HSP of $dH/dt$ for ARs (red diamonds without error bars) in Figure~\ref{fig:hsp_1_ar} are also shown for comparison.}
\label{fig:hsp_1_pr}
\end{figure}

\begin{figure}[t!]
\centering
\includegraphics[width=0.78\textwidth]{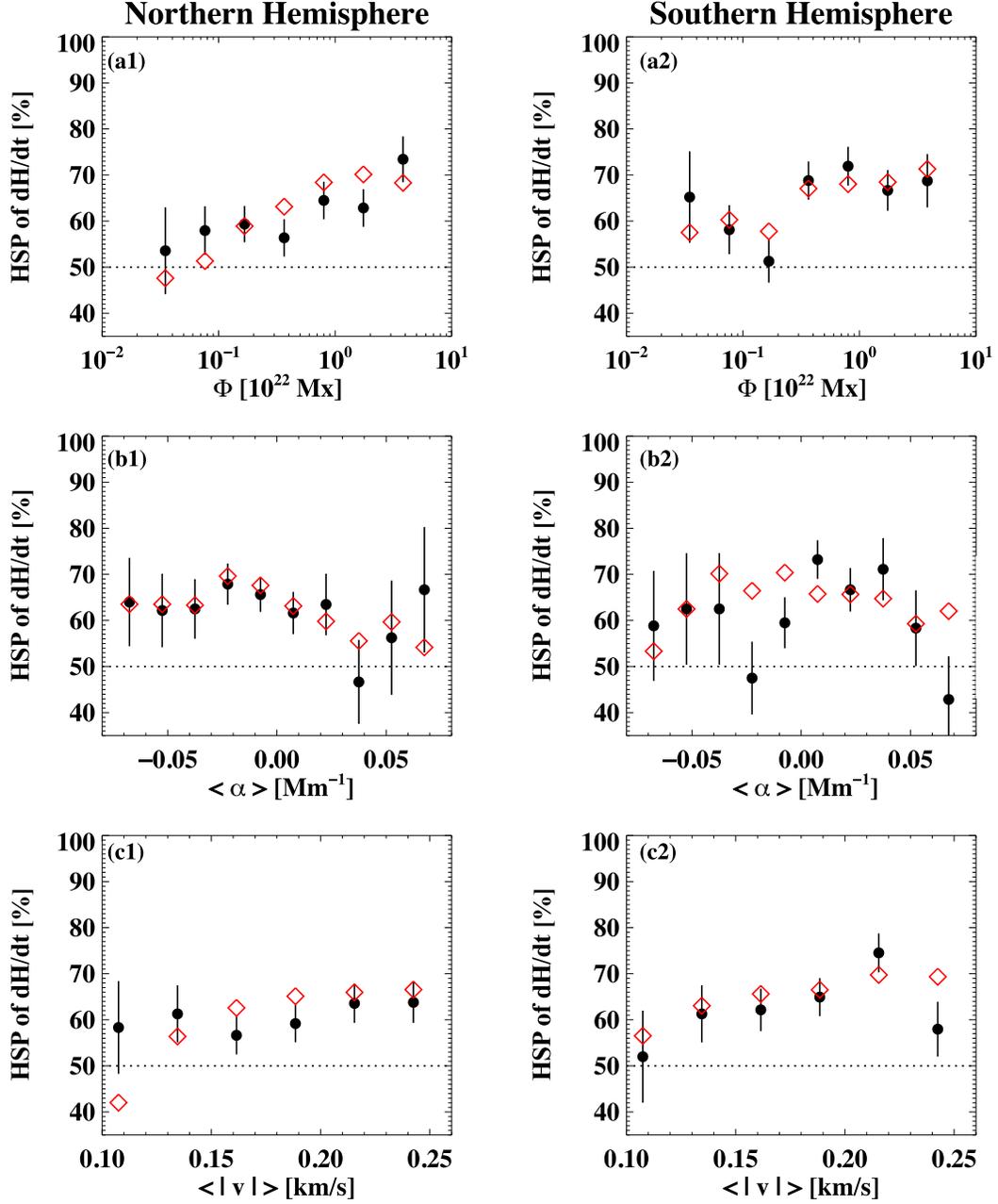}
\caption{Same as ARs (red diamonds) shown in Figure~\ref{fig:hsp_2_ar}, but for plage regions (circles with error bars). The HSP of $dH/dt$ for ARs (red diamonds without error bars) in Figure~\ref{fig:hsp_2_ar} are also shown for comparison.}
\label{fig:hsp_2_pr}
\end{figure}

Plage regions are observed as bright areas in H$\alpha$ line center images without a dark sunspot in the continuum. In particular with the dataset considered here, plage regions may be assumed to be aged regions with decaying magnetic fields. In this section we investigate whether there is any difference of the HSP of $dH/dt$ between ARs and plages. For this, we first searched HARPs over the same investigated period of $\sim$8 years (i.e., 2010 May 1 to 2017 December 31), each of which contains only a single NOAA-numbered plage identified in the NOAA Solar Region Summary archive (\url{ftp://ftp.swpc.noaa.gov/pub/warehouse}). A total of 1,456 daily samples are assigned to 488 unique NOAA-identified plage areas, specifically 795 and 661 samples in the northern and southern hemispheres, respectively. Figures~\ref{fig:hsp_1_pr} and~\ref{fig:hsp_2_pr} show the HSP of $dH/dt$ for plage samples (black circles with error bars) as a function of the region properties described in Figures~\ref{fig:hsp_1_ar} and~\ref{fig:hsp_2_ar}, with the HSP of $dH/dt$ for ARs (red diamonds) also shown for reference. We find that ARs and plages exhibit similar overall trends in the strength of the HSP of $dH/dt$ with respect to the properties considered (explained in detail in Section~\ref{subsec:results_hsp}). Even though some differences are found in the HSP trends between ARs and plages (e.g., see panel (c1) of Figure~\ref{fig:hsp_2_pr} for samples with $<\!\!|v|\!\!>$ below 0.2 km/s), it should be noted that the statistical significance of such differences is limited, due to the comparatively small number of plage samples.

\end{document}